\documentclass[submission,copyright,creativecommons]{eptcs}
\providecommand{\event}{MARS 2020} 

\usepackage[utf8]{inputenc}
\usepackage{url}
\usepackage{amsmath,amssymb,amsfonts}
\usepackage{amsthm}
\usepackage{enumitem} 
\usepackage{braket}
\usepackage{siunitx}
\usepackage{booktabs}
\usepackage{wrapfig}
\usepackage{tikz}
\usepackage{lineno} 
\usepackage{array}

\usetikzlibrary{arrows,shapes,snakes,calc,automata,patterns}

\newcolumntype{L}[1]{>{\raggedright\let\newline\\\arraybackslash\hspace{0pt}}m{#1}}
\newcolumntype{C}[1]{>{\centering\let\newline\\\arraybackslash\hspace{0pt}}m{#1}}
\newcolumntype{R}[1]{>{\raggedleft\let\newline\\\arraybackslash\hspace{0pt}}m{#1}}

\definecolor{fg0}{RGB}{33,113,181}
\definecolor{fg0medium}{RGB}{116,169,207}
\definecolor{fg0light}{RGB}{189,215,231}
\definecolor{fg1}{RGB}{49,163,84}
\definecolor{fg2}{RGB}{49,163,84}
\definecolor{hghlght}{RGB}{127,205,187}

\usepackage[textsize=scriptsize]{todonotes}

\DeclareMathOperator{\tOccMin}{occ\_min}
\DeclareMathOperator{\tOccMax}{occ\_max}

\DeclareMathOperator{\tLength}{length}

\DeclareMathOperator{\tSignals}{sigs}
\DeclareMathOperator{\tPDU}{PDU}
\DeclareMathOperator{\tFrame}{frame}

\DeclareMathOperator{\TimingModel}{\mathbf{T\mkern-1mu M}}
\DeclareMathOperator{\TimingModelP}{\mathbf{T\mkern-1mu M\mkern-1mu P}}
\DeclareMathOperator{\TimingModelS}{\mathbf{T\mkern-1mu M\mkern-1mu S}}

\let\mod\relax
\DeclareMathOperator{\mod}{mod}

\newtheorem{definition}{Definition}

\title{Estimating End-to-End Latencies in Automotive Cyber-physical Systems}

\author{Max J. Friese
	\institute{Mercedes-Benz AG\\ Sindelfingen, Germany}
	\institute{Department of Computer Science\\
		Kiel University\\
		Kiel, Germany}
	\email{max\_jonas.friese@daimler.com}
	\and
	Dirk Nowotka
	\institute{Department of Computer Science\\
		Kiel University\\
		Kiel, Germany}
	\email{dn@informatik.uni-kiel.de}
}

\def\authorrunning{M.J. Friese \& D. Nowotka}
\def\titlerunning{Estimating End-to-End Latencies in Automotive Cyber-physical Systems}

\begin{document}
	
	\maketitle
	
	\begin{abstract}
		Controller networks in today’s automotive systems consist of more than 100 ECUs 
		connected by various bus protocols. 
		Seamless operation of the entire system requires a well-orchestrated interaction 
		of these ECUs. 
		Consequently, to ensure safety and comfort, a performance analysis is an inherent part 
		of the engineering process. 
		Conducting such an analysis manually is expensive, slow, and error prone. 
		Tool support is therefore crucial, and a number of approaches have been presented. 
		However, most work is limited to either network latencies or software latencies which 
		results in an analysis gap at the transition between different layers of the 
		communication stack. The work presented here introduces an approach to close this gap. 
		Furthermore, we discuss the integration of different methods to obtain an end-to-end latency analysis.
	\end{abstract}
		
	\section{Introduction}
	Distributed cyber-physical systems (CPS) are the key for many
	technology developments in today's connected world, 
	one prominent example is the development of automated vehicles.
    Advancing \emph{classic} embedded computing, CPS stand 
    for the integration of computing and  physical processes in networks of 
    heterogeneous components. Due to their distributed nature, system 
    functions span over multiple devices and possibly contain feedback loops \cite{EALee06}. 
	Two aspects need to be considered to assess the correct behavior 
	of a CPS with regards to a system function: \emph{how} does the system 
	react and \emph{when} does the system react. More precisely, if a certain 
	stimulus occurs, the system has to react correctly and at the 
	correct time, e.g. an automated car should start to brake immediately, 
	if its sensors detect a pedestrian on the road. Consequently, the time 
	to react is part of the requirements for the correct behavior. 

	Accordingly, the specification and validation of the temporal behavior 
	of CPS is an important part of system's engineering. The latter is getting
	considerably more complex due to the continuing growth of hard- and software 
	architectures. The former usually starts with an end-to-end time budget for 
	a system function, e.g. the backup camera should show an image within 
	 $\SI{300}{ms}$ after the car was put into reverse. 	
	To verify that the system meets the timing constraints for a function, one 
	or more cause-effect chains are analyzed. 
	Each cause-effect chain describes one flow of data 
	through the system in detail. 
	The cause-effect chains are further subdivided and broken down 
	into smaller parts, e.g. the sub-chain within one 
	electronic control unit (ECU).
	Each sub-chain contributes to the end-to-end latency and 
	receives a share of the timing budget accordingly.
	The implementer of the different sub-chains have to assure that their 
	part of the chain meets its timing constraints. 
	
	The latency caused within the \emph{cyber} part of the system 
	falls into one of the following categories: 
	(1) latencies due to processing of software, or
	(2) latencies due to network communication. 
	For the software part, 
	the possible core execution times of the involved tasks and 
	the dataflow through the tasks is analyzed to find the path which yields 
	the worst-case latency. 
	The latency induced by network communication 
	is the sum of the so-called transmission and the so-called 
	propagation delay. In summary: the analysis of software 
	latencies ends at the last write access on some variable
	in global memory and the analysis of network latencies starts 
	at the transmission of a network frame. 
	
	In this work we argue that meeting local timing constraints 
	is no sufficiency for meeting end-to-end timing constraints 
	if partitioning of the chain is not done 
	carefully. In this context we 
	call attention to an important aspect currently not sufficiently 
	considered in the integration of software- and network analyses. 
	It originates in the need for increased bandwidth which lead to 
	a relaxation in the static mapping of protocol data units (PDUs)  
	into their encapsulating PDUs. 
	To increase flexibility and thereby save network resources, 
	e.g. by sending two encapsulated PDUs alternately, 
	PDU mapping is done dynamically. 
	The actual mapping is determined at runtime. 
	On the data link layer, PDUs are triggered for 
	sending either when full, when certain timers expire, or 
	when a higher-layer PDU needs to be sent immediately. 
	The combination of event-based packing, immediate sending, 
	and dynamic mapping leads to complex situations 
	where it is not directly evident whether an updated value 
	is sent with the very next encapsulating PDU. The impacts 
	of these mechanisms on end-to-end timings are currently 
	neither considered in software-focused nor in network-focused
	analyses, leaving a gap in the methodology for formally 
	derived, safe end-to-end estimations. 
	
	\subsection{Contribution}
	The contributions of this paper are twofold. First, we present an
	analysis model and technique for the temporal behavior of  
	PDU-transmission mechanisms, which are commonly used in automotive 
	CPS. 
	Secondly, we discuss the integration of formal methods to obtain  
	end-to-end estimations. In particular we 
	\begin{itemize}
		\item introduce \emph{timing models} and show how two 
		      or more can be combined to model the composite 
		      of temporal behaviors. 
		\item present an approach to encode different trigger mechanisms 
		      for signals, PDUs and frames to obtain safe estimations on their 
		      impact on end-to-end latencies. 
		\item describe how to integrate this approach with analyses for 
			  software-level estimations to obtain end-to-end latencies. 
		\item report the applicability on an industrial-scale use case.
		        
	\end{itemize}
	
	\subsection{Outline}
	This paper is organized as follows: related work is presented 
	in the next section. Subsequently, we describe the systems 
	considered here in Section~\ref{sec:SystemModel}. 
	In Section~\ref{sec:EstimationsNw} we present our approach 
	to estimate the impact of different PDU triggering mechanisms 
	on end-to-end timing. We use CAN-FD as an example for the 
	data link layer and show how to plug-in respective analyses. 
	In Section~\ref{sec:EstmationsE2E} we discuss how to integrate 
	the approach of the previous section with existing approaches 
	for the software part to obtain an end-to-end latency. 
	To demonstrate applicability, we report the application of  
	the approach on an automotive use case in Section~\ref{sec:Experiments}.	
	Finally, in Section~\ref{sec:FutureWork} and Section~\ref{sec:Conclusion}
	we provide an insight into planned future work and conclude. 
	
	\section{Related Work} \label{sec:RelatedWork}	
	Related work comes from two categories. 
	Firstly, these are approaches to estimate the
	latency due to network communication, i.e. the  
	delay between a network controller of one 
	ECU and the network controller of the next ECU 
	in the chain. Respective analyses have been presented 
	for different automotive field busses 
	\cite{DBLP:journals/rts/DavisBBL07,DBLP:conf/sies/NeukirchnerNEB12}.
    So-called \emph{holistic} approaches are additionally 
    concerned with parts of the ECU's software \cite{DBLP:conf/iecon/LangeBVO16}. 
    However, currently they rely on single-core analyses \cite{DBLP:journals/csi/LangeOV16}. 
	Earlier, so-called \emph{compositional} approaches were 
	developed to cope complexity and therefore make analysis 
	applicable. In compositional performance analysis (CPA)
	different local scheduling analyses are combined 
	to obtain end-to-end estimations \cite{DBLP:conf/dac/RichterZJE02}. 
	CPA got a lot of research attention, and was used e.g. 
	for the end-to-end response time analysis in automotive systems 
	\cite{DBLP:journals/tcad/SchlieckerRNRJE09}. 
	However, in its basic form, CPA suffers from the problem that 
	multiple worst-cases are possibly considered simultaneously 
	although they can not occur at the same time.  
	Recent work in the area underlines applicability for
	industrial-scale use cases 
	\cite{gemlau2017compositional,DBLP:journals/rts/ThieleSAE16}
	and improves, i.e. reduces, pessimism of the estimations 
	\cite{DBLP:conf/date/KohlerNEB19}.
	
	Related work from the second category is concerned with 
	estimations for the time needed to propagate signal changes 
	through the software within an ECUs. For automotive CPS, 
	this analysis is performed by analyzing the end-to-end response 
	time of task chains. 
	Two types of task chains are distinguished. 
	On the one hand, there are chains where tasks 
	activate their successors based on events generated by changes 
	in signal values \cite{DBLP:conf/rtas/SchlatowE16}.
	These chains are also referred to as \emph{functional chains} 
	\cite{DBLP:journals/tcad/GiraultPQHS18}. 
	On the other hand, there are chains through tasks activated by 
	periodic timers; these are so-called \emph{cause-effect chains}.
	For the sake of simplicity, we will refer to both kinds 
	of chains as cause-effect chains in this work. 	
	Two latency semantics are of interest for cause-effect chains: 
	\emph{response time} is the time needed to react 
	to a certain input, and \emph{data age} is the time span an 
	input has still an impact on the output \cite{Feiertag2008}. 
	Like functional chains, cause-effect chains in software are well-studied 
	in terms of algorithmic approaches to estimate latencies 
	\cite{DBLP:conf/rtcsa/BeckerDMBN16,DBLP:journals/jsa/BeckerDMBN17}.		
	Other approaches to tackle this problem include 
	encoding of the system's behavior in integer linear programming 
    \cite{DBLP:conf/nfm/BoniolLPE13,DBLP:journals/ijccbs/LauerBPE14}
    and constraint programming \cite{DBLP:conf/sies/FrieseEN18}. 
	Furthermore, the work around the logical 
	execution time paradigm (LET) \cite{DBLP:conf/emsoft/HenzingerHK01} 
	must be attributed, as it aims to simplify timing analysis through 
	deterministic behavior, also on system-level \cite{DBLP:conf/iecon/ErnstAG18}.
	Corresponding work was done in researching LETs 
	potential to determine end-to-end latencies
	\cite{DBLP:conf/ecrts/HamannD0PW17,DBLP:journals/tcad/MartinezOB18}. 

	However, integrating models from different development stages, 
	and therefore with different levels of details to determine 
	end-to-end latencies in all stages of systems engineering is still 
	an ongoing challenge. The constraint programming (CP) approach followed 
	in this paper is especially promising because the level of detail 
	can easily adjusted by adding or removing constraints. 
	
	\section{System Model}\label{sec:SystemModel}
	Network clusters in modern automotive systems are usually divided into
	multiple domains focused on a set of functionalities of the
	car, e.g. powertrain or infotainment \cite{Jiang2019}. 
	Each domain comprises a set of controllers collaborating 
	to implement system functions. 	
	Additionally, gateway controllers
	encapsulate the communication with other domains. 
	The topology of the network is also reflected in the  
	communication design. Communication can be divided into two categories. 
	On the one hand there are classic 
	field busses like LIN, CAN or FlexRay which are compatible 
	with static resource allocation of real-time embedded systems. 
	In current architectures they are commonly 
	used to interconnect ECUs within a domain. 
	Due the increasing amount of bandwidth-intense applications, 
	like e.g. image recognition, domain controllers,
	on the other hand, are connected via Ethernet backbones. 
	The Ethernet deployment coincides with a paradigm shift from 
	signal-based to service-oriented communication, especially for 
	inter-domain communication. In signal-based communication 
	the possible data exchange between ECUs is static and concluded 
	after design, whereas service-oriented communication 
	allows to extend communication dynamically. 
	However, both paradigms have in common that PDUs are used 
	for the vertical communication through the stack within an ECU.
	Different mechanisms exists to trigger the transmission of PDUs. 
	After a PDU was triggered, e.g. by a timeout, it is encapsulated 
	in a lower-layer PDU. Once the data link layer is reached, 
    PDUs are put on the physical medium by the communication 
    controller of the ECU. 
    Nowadays, for some PDUs, encapsulation is decided 
    dynamically at runtime. As a consequence, an update in a signal 
    value is not necessarily sent 
	in the next available PDU if another PDU is first in line.
	The postponed PDU is sent with a subsequent transmission.
	This leads to complex scenarios aggravating a manual 
	verification of the end-to-end latency constraints 
	of distributed system functions. 
	
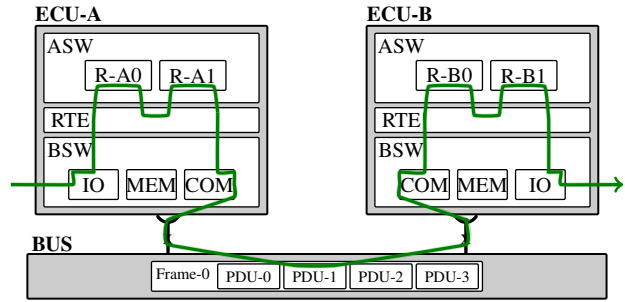
\begin{figure}[t]
		\centering\scriptsize
		
		\begin{tikzpicture}[xscale=1.10,yscale=0.66] 
		
		\tikzstyle{ecu} = [line width=0.75pt, color=black, fill=black!20!white, rounded corners=0.5pt]
		\tikzstyle{con} = [line width=1.25pt, color=black, fill=black!20!white ]
		\tikzstyle{bus} = [line width=0.75pt, color=black, fill=black!20!white, rounded corners=0.5pt]
		\tikzstyle{sw}  = [line width=0.50pt, color=black, fill=white, rounded corners=0.5pt]
		\tikzstyle{art} = [line width=0.50pt, color=black, fill=white, rounded corners=0.5pt]
		
		\tikzstyle{chain} = [color=green!50!black, line width=1.8pt, smooth, opacity=0.25]
		\tikzstyle{chainA} = [color=green!50!black, line width=1.2pt, smooth]
		\tikzstyle{chainB} = [densely dotted, color=green!40!black, line width=1.2pt, smooth]
		\tikzstyle{chainC} = [->, color=green!50!black, line width=1.2pt, smooth]
		
		\foreach \o / \n in { 0.0/A, 4.0/B} { 
			\draw[ecu] (\o+ 0.00,-0.05) rectangle ++ ( 2.80, 3.80) ; \node[draw=none] at (\o+ 0.40, 4.00) {\textbf{ECU-\n}};
			\draw[sw ] (\o+ 0.10, 0.10) rectangle ++ ( 2.60, 1.40) ; \node[draw=none] at (\o+ 0.42, 1.25) {BSW};
			\draw[sw ] (\o+ 0.10, 1.60) rectangle ++ ( 2.60, 0.50) ; \node[draw=none] at (\o+ 0.42, 1.85) {RTE};
			\draw[sw ] (\o+ 0.10, 2.20) rectangle ++ ( 2.60, 1.40) ; \node[draw=none] at (\o+ 0.42, 3.35) {ASW};
			
			\if\n A	
			\draw[art] (\o+ 0.40, 0.25) rectangle ++ ( 0.60, 0.60) node[draw=none,pos=.5] (IOA)  {IO};
			\draw[art] (\o+ 1.10, 0.25) rectangle ++ ( 0.60, 0.60) node[draw=none,pos=.5] (MEMA) {MEM};
			\draw[art] (\o+ 1.80, 0.25) rectangle ++ ( 0.60, 0.60) node[draw=none,pos=.5] (COMA) {COM};
			
			\draw[con, )-] (\o+ 1.60,-0.06) -- ++( 0.00,-1.00) node[draw=none,pos=.5] (CON\n) {x} ;
			\else
			\draw[art] (\o+ 0.40, 0.25) rectangle ++ ( 0.60, 0.60) node[draw=none,pos=.5] (COMB) {COM};
			\draw[art] (\o+ 1.10, 0.25) rectangle ++ ( 0.60, 0.60) node[draw=none,pos=.5] (MEMB) {MEM};
			\draw[art] (\o+ 1.80, 0.25) rectangle ++ ( 0.60, 0.60) node[draw=none,pos=.5] (IOB)  {IO};
			
			\draw[con, )-] (\o+ 1.20,-0.06) -- ++( 0.00,-1.00) node[draw=none,pos=.5] (CON\n) {x} ;
			\fi 
			
			\draw[art] (\o+ 0.60, 2.45) rectangle ++ ( 0.80, 0.60) node[draw=none,pos=.5] (R\n0) {R-\n0};
			\draw[art] (\o+ 1.50, 2.45) rectangle ++ ( 0.80, 0.60) node[draw=none,pos=.5] (R\n1) {R-\n1};
			
		}
		
		\node[draw=none] at (0.20, -0.65) {\textbf{BUS}};
		\draw[bus] (-0.10,-1.75) rectangle ++ ( 7.00, 0.90) ; 
		\draw[art]       ( 1.40,-1.60) rectangle ++ ( 4.00, 0.60); 
		\node[]       at ( 1.78,-1.25) {\tiny Frame-0}; 
		\draw[art]       ( 2.20,-1.55) rectangle ++ ( 0.75, 0.50) node[draw=none,pos=.5] (PDU0) {\tiny PDU-0}; 
		\draw[art]       ( 3.00,-1.55) rectangle ++ ( 0.75, 0.50) node[draw=none,pos=.5] (PDU1) {\tiny PDU-1}; 
		\draw[art]       ( 3.80,-1.55) rectangle ++ ( 0.75, 0.50) node[draw=none,pos=.5] (PDU2) {\tiny PDU-2}; 
		\draw[art]       ( 4.60,-1.55) rectangle ++ ( 0.75, 0.50) node[draw=none,pos=.5] (PDU3) {\tiny PDU-3}; 
		
		\draw [chain] plot [smooth, tension=.15] coordinates { 
			($ (IOA)  + (-1.00, 0.00) $)  	 
			($ (IOA)  + (-0.25, 0.00) $)     
			($ (IOA)  + (-0.25, 0.25) $)     
			($ (IOA)  + ( 0.00, 0.25) $)       
			($ (IOA)  + ( 0.00, 0.30) $)        
			($ (RA0)  + (-0.30,-0.90) $)           
			($ (RA0)  + (-0.25,-0.22) $)                
			($ (RA0)  + ( 0.25,-0.22) $) 
			($ (RA0)  + ( 0.30,-0.80) $) 
			($ (RA1)  + (-0.30,-0.80) $) 
			($ (RA1)  + (-0.25,-0.22) $)  
			($ (RA1)  + ( 0.25,-0.22) $)  
			($ (RA1)  + ( 0.30,-0.75) $) 
			($ (COMA) + ( 0.10, 0.25) $) 
			($ (COMA) + ( 0.30, 0.25) $)
			($ (COMA) + ( 0.30, 0.00) $) 
			($ (COMA) + ( 0.30,-0.25) $) 
			($ (CONA) + ( 0.00, 0.40) $) 
			($ (CONA) + ( 0.00,-0.10) $) 
			($ (PDU1) + (-0.25, 0.22) $) 
			($ (PDU1) + ( 0.25, 0.22) $) 
			($ (CONB) + ( 0.00,-0.10) $) 
			($ (CONB) + ( 0.00, 0.40) $) 
			($ (COMB) + (-0.30,-0.25) $) 
			($ (COMB) + (-0.30, 0.00) $) 
			($ (COMB) + (-0.30, 0.25) $) 
			($ (COMB) + ( 0.00, 0.25) $)
			($ (RB0)  + (-0.30,-0.90) $)           
			($ (RB0)  + (-0.25,-0.22) $)                
			($ (RB0)  + ( 0.25,-0.22) $) 
			($ (RB0)  + ( 0.30,-0.80) $)
			($ (RB1)  + (-0.30,-0.80) $) 
			($ (RB1)  + (-0.25,-0.22) $)  
			($ (RB1)  + ( 0.25,-0.22) $)  
			($ (RB1)  + ( 0.30,-0.90) $)
			($ (IOB)  + ( 0.10, 0.30) $)     
			($ (IOB)  + ( 0.10, 0.25) $)     
			($ (IOB)  + ( 0.25, 0.25) $)
			($ (IOB)  + ( 0.25, 0.00) $)
			($ (IOB)  + ( 1.00, 0.00) $) } ;
		
		\draw [chainA] plot [smooth, tension=.15] coordinates { 
			($ (IOA)  + (-1.00, 0.00) $)  	 
			($ (IOA)  + (-0.25, 0.00) $)     
			($ (IOA)  + (-0.25, 0.25) $)     
			($ (IOA)  + ( 0.00, 0.25) $)       
			($ (IOA)  + ( 0.00, 0.30) $)        
			($ (RA0)  + (-0.30,-0.90) $)           
			($ (RA0)  + (-0.25,-0.22) $)                
			($ (RA0)  + ( 0.25,-0.22) $) 
			($ (RA0)  + ( 0.30,-0.80) $) 
			($ (RA1)  + (-0.30,-0.80) $) 
			($ (RA1)  + (-0.25,-0.22) $)  
			($ (RA1)  + ( 0.25,-0.22) $)  
			($ (RA1)  + ( 0.30,-0.75) $)  }; 
		
		\draw [chainA] plot [smooth, tension=.15] coordinates {    
			($ (RA1)  + ( 0.30,-0.75) $) 
			($ (COMA) + ( 0.10, 0.25) $) 
			($ (COMA) + ( 0.30, 0.25) $)
			($ (COMA) + ( 0.30, 0.00) $) 
			($ (COMA) + ( 0.30,-0.25) $) 
			($ (CONA) + ( 0.00, 0.40) $) 
			($ (CONA) + ( 0.00,-0.10) $) 
			($ (PDU1) + (-0.25, 0.22) $) 
			($ (PDU1) + ( 0.25, 0.22) $) 
			($ (CONB) + ( 0.00,-0.10) $) 
			($ (CONB) + ( 0.00, 0.40) $) 
			($ (CONB) + ( 0.00, 0.40) $) 
			($ (COMB) + (-0.30,-0.25) $) 
			($ (COMB) + (-0.30, 0.00) $) 
			($ (COMB) + (-0.30, 0.25) $) 
			($ (COMB) + ( 0.00, 0.25) $)
			($ (RB0)  + (-0.30,-0.75) $) };
		
		\draw [chainC] plot [smooth, tension=.15] coordinates { 
			($ (RB0)  + (-0.30,-0.90) $)           
			($ (RB0)  + (-0.25,-0.22) $)                
			($ (RB0)  + ( 0.25,-0.22) $) 
			($ (RB0)  + ( 0.30,-0.80) $)
			($ (RB1)  + (-0.30,-0.80) $) 
			($ (RB1)  + (-0.25,-0.22) $)  
			($ (RB1)  + ( 0.25,-0.22) $)  
			($ (RB1)  + ( 0.30,-0.90) $)
			($ (IOB)  + ( 0.10, 0.30) $)     
			($ (IOB)  + ( 0.10, 0.25) $)     
			($ (IOB)  + ( 0.25, 0.25) $)
			($ (IOB)  + ( 0.25, 0.00) $)
			($ (IOB)  + ( 1.00, 0.00) $) } ;
		\end{tikzpicture}
		
		\caption{Cause-effect Chain (green) spanning over two ECUs}
		\label{fig:E2EEstimations}
	\end{figure}

	Within an ECUs the data flow for computing a system 
	function passes multiple so-called runnables as depicted 
	in Figure~\ref{fig:E2EEstimations} (e.g. R-A0). 
	A runnable is a piece of software dedicated to the 
	computation to parts of the overall functionality. 
	For processing, multiple runnables are combined to 
	a task. Task sets of automotive ECUs 
	comprise multiple, periodically activated 
	tasks with different rates and offsets. Additionally, 
	sporadically activated task handle interrupts, e.g. 
	for sensor readings. The signals used to exchange values 
	are stored in variables located in a global memory. 
	
	At the bound of the ECU and the network, dedicated 
	communication tasks copy data from global 
	memory to the buffer of the communication controller 
	if transmission requested.
	In this work, we divide the cause-effect chains 
	associated to a time-constrained system function 
	into two kinds of sub-chains: (1) communication task 
	to communication task via network and (2)
	communication task to communication task via task chain. 
	The course of events for both kind of chain 
	is mainly driven the possible time intervals 
	memory is accessed by the runnables of a task. 
	In this work, we consider implicit data 
	reception and transmission on task-level \cite{Autosar43/RTE}, 
	meaning that a task receives all signal values when 
	it is activated and that all signal values are 
	written collectively when the task terminates. 
	
	We describe how to obtain estimations for the 
	first kind of sub-chain in Section~\ref{sec:EstimationsNw}. 
	The problem of estimating latencies for task chains 
	is well researched. We briefly revisit the problem in 
	Section~\ref{sec:EstimationsSW} in order to discuss 
	how to obtain \emph{overall} end-to-end latencies in  
	Section~\ref{sec:EstmationsE2E}. 	

	\section{Formal abstractions} \label{sec:FormalAbstractions}
	The temporal behavior of the system with regard to a cause-effect 
	chain depends on the possible time intervals for different events, 
	e.g. a PDU being triggered for sending. The trigger mechanism 
	for PDUs are described in detail in the next section. 
	However, less specifically speaking, a PDU can be triggered 
	by a timer or due to a value change in a signal. 
	As described above, the latter depends on the points 
	of time a task produces an update for the signal value. 
	Although signals are likely to be produced by a 
	dedicated runnable and therefore written by a 
	single task, in general it is possible that multiple 
	tasks write a signal. To describe and combine 
	intervals of time in which this might happen, 
	we introduce the notion of \emph{Timing Models}. 
	Our formalization is based on a discrete and finite time model: 
	let 
	$\mathbb{T} = \set{T_\text{min},\dots,T_\text{max}} \subseteq \mathbb{N}$ 
	be the time domain with $T_\text{min}$ and $T_\text{max}$
	as minimum and maximum valid points of time. 
	Furthermore, let $T_\text{sup} = \sup\left(\mathbb{T}\right)$ be 
	a value to indicate invalid points of time. 	
	
	\subsection{Timing Models} 
	To model the temporal behavior of the system that is 
	being analyzed, we consider different events, like e.g. 
	a PDU being triggered for sending. 
	Similar to the \emph{Arrival Curves} introduced in \cite{DBLP:journals/tit/Cruz91} 
	we use \emph{Timing Models} to describe the 
	nature of how events occur. However, unlike the 
	\emph{Cumulative Function} of \emph{Network Calculus} \cite{DBLP:books/sp/BoudecT01}
	or the 
	\emph{Interval Bound Functions} of \emph{Real-time Calculus} \cite{DBLP:conf/iscas/ThieleCN00}
	we do not use them to derive request and response counts 
	but use them to bound the interval of time in which an 
	event might occur within $\mathbb{T}$. 
		
	\begin{definition}[Timing Model]\label{def:TimingModel}
		A timing model is a function $m:\mathbb{N}\rightarrow\mathbb{N}\times\mathbb{N}$ 
		which maps the occurences of an event to 
		the first and the last possible point of time the event might occur. 
		For elements of the codomain we use the projection functions 
		$\pi_1$ and $\pi_2$ to access the respective element, 
		e.g. let $p=(a,b)\in\mathbb{N}\times\mathbb{N}$, then 
		$\pi_1(p)=a$ and $\pi_2(p)=b$.  
	\end{definition}
	
	We distinguish two types of timing models. Firstly, 
	\emph{Periodic Timing Models} are used to describe events 
	triggered by periodic clocks. The time in which these events 
	might occur does not vary. Secondly, \emph{Sporadic Timing Models}
	allow for the specification of temporally less predetermined events. 
	Here, the time span for an occurrence of the event can only be 
	be narrowed down to a minimum and maximum interarrival time. 
	
	\begin{definition}[Periodic Timing Models]\label{def:PeriodicTM}
		A periodic timing model $t^\mathbf{P}_{o,p,n}:\mathbb{N}\rightarrow\mathbb{N}\times\mathbb{N}$ 
		is a timing model parameterized with three arguments
		$o\in\mathbb{N}$ and $p,n\in\mathbb{N}_{>0}$
		with $t^\mathbf{P}_{o,p,n} (i) = (a_i, a_i)$ 
		where $a_i = o + \left\lceil \tfrac{i}{n}\right\rceil \cdot p$
		for all $i\in\mathbb{N}$. 
		The family of periodic timing models
		is defined as $\TimingModelP = \set{t_{o,p,n} | o\in\mathbb{N}, p,n\in\mathbb{N}_{>0}}$. 
	\end{definition}

	\begin{definition}[Sporadic Timing Models]\label{def:SporadicTM}
		A sporadic timing model $t^\mathbf{S}_{l,u,n}:\mathbb{N}\rightarrow\mathbb{N}\times\mathbb{N}$ 
		is a timing model parameterized with three arguments 
		$l,n\in\mathbb{N}_{>0}$ and $u\in\mathbb{N}_{>l}$
		with $t^\mathbf{S}_{l,u,o} (i) = (l_i, u_i)$ where  
		\begin{align*}
		l_i = \left\lceil \tfrac{i}{n} \right\rceil \cdot l \;\;\text{and}\;\; 
		u_i = \left(\left\lceil \tfrac{i}{n} \right\rceil + 1\right) \cdot u
		\end{align*} 
		for all $i\in\mathbb{N}$. 
		The family sporadic timing models
		is defined as $\TimingModelS = \set{t^\mathbf{S}_{l,u,n} | l,n\in\mathbb{N}_{>0}, u\in\mathbb{N}_{>l}}$
	\end{definition}

	Let $\TimingModel = \TimingModelP \cup \TimingModelS$ be the 
	set of timing models.
	Finally, to combine multiple possible time intervals for an 
	event, we define the $\sqcup$-operator for timing models. 

	\begin{definition}[Union of Timing Models]\label{def:TMArithmetics}
		Let $t_0, t_1\in\TimingModel$ be timing models. To define the union 
		of $t_0$ and $t_1$, formally $t_0\sqcup t_1$, we distinguish three 
		cases: 
		\begin{description}[style=unboxed,leftmargin=0cm]
			\item[\textbf{Case~1}] 
			$t_0=t^\mathbf{P}_{o_0,p_0,n_0}$ is a periodic timing model and 
			$t_1=t^\mathbf{P}_{o_1,p_1,n_1}$ is a periodic timing model. 	                                          
			\begin{align*}
			t_{o_0,p_0} \sqcup t_{o_1,p_1} = 
			\begin{cases}
			t^\mathbf{P}_{o,p',n'} 
			&\text{ if } o_0 = o_1 \land \mod\left(p_\text{max},p_\text{min}\right)= 0  \\
			t^\mathbf{S}_{l,u,n'} 
			&\text{ else }  
			\end{cases}     
			\end{align*}
			with 		
			$p_\text{min} = \min\set{p_0,p_1}$,
			$p_\text{max} = \max\set{p_0,p_1}$,
			$l = \min\set{p_0,p_1}$ ,
			$u = \max\set{o_0,o_1}+p_\text{max}$,
			$n' = n_0 + n_1$, and 
			$p' = p_\text{min}$.			
			
			\item[\textbf{Case~2}] 
			$t_0=t^\mathbf{S}_{l,u,n_0}$ is a sporadic timing model and 
			$t_1=t^\mathbf{P}_{o,p,n_1}$ is a periodic timing model. 
			\begin{align*}
			t^\mathbf{S}_{l,u,n_0} \sqcup t^\mathbf{P}_{o,p,n_1}
			= t^\mathbf{P}_{o,p,n_1} \sqcup t^\mathbf{S}_{l,u,n_0}
			= t^\mathbf{S}_{l',u',n'}
			\end{align*}
			with 
			$l'=\min\set{l,p}$,
			$u'=\max\set{o+p,u}$, and 
			$n'=n_0 + n_1$. 
			
			\item[\textbf{Case~3}]
			$t_0=t^\mathbf{S}_{l_0,u_0,n_0}$ is a sporadic timing model and 
			$t_1=t^\mathbf{S}_{l_1,u_1,n_1}$ is a sporadic timing model. 
			\begin{align*}
			t^\mathbf{S}_{l_0,u_0,n_0} \sqcup t^\mathbf{S}_{l_1,u_1,n_1} 
			= t^\mathbf{S}_{l',u',n'}
			\end{align*}
			with 
			$l' = \min(\set{l_0,l_1,\vert l_0 - l_1\vert})$, 
			$u' = \max(\set{u_0,u_1})$, and 
			$n' = n_0 + n_1$. 
		\end{description}
				
	\end{definition}

	The $\sqcup$-operator allows us to combine arbitrary 
	timing models for events.  This is particularly suitable for
	describing all possible points of time at which a signal value 
	might be updated. Consider a signal written by two 
	runnables processed in two different tasks. One task 
	is activated every $\SI{5}{ms}$ and without an offset 
	and the other one is activated every $\SI{10}{ms}$ 
	with an offset of $\SI{2.5}{ms}$. This means, updates of the 
	signal will happen after the following points of time: \\
	$
	 \qquad \text{first:}   \SI{0}{ms} \qquad \text{second:}   \SI{2.5}{ms} \qquad \text{third:}   \SI{5}{ms} 
	 \qquad \text{fourth:} \SI{10}{ms} \qquad \text{fifth:}   \SI{12.5}{ms} \qquad \text{sixth:}  \SI{15}{ms} 
	$. 
	
	This pattern is safely approximated by the sporadic activation pattern 
	$t^\mathbf{S}_{5000,12500,2}$ which 
	gives the following  possible intervals for 
	the $i^\text{th}$ update: 
	\begin{align*}
	\begin{array}{lll}
	 t^\mathbf{S}_{5000,12500,2}(1) = (    0,12500) &  
	 t^\mathbf{S}_{5000,12500,2}(2) = (    0,12500) &
	 t^\mathbf{S}_{5000,12500,2}(3) = ( 5000,25000) \\
	 t^\mathbf{S}_{5000,12500,2}(4) = ( 5000,25000) &
	 t^\mathbf{S}_{5000,12500,2}(5) = (10000,37500) &  
	 t^\mathbf{S}_{5000,12500,2}(6) = (10000,37500) 
	\end{array}
	\end{align*}
	
	Safely here means, that the possible point of time for the $n^\text{th}$
	activation lies within the value of the timing model for $n$.  
	In the next section we use this to describe the complex temporal behavior 
	of different PDU triggers based on the potential updates of signal values. 

	\subsection{Estimations on Network-level} \label{sec:EstimationsNw}	
    Based on the system model described in Section~\ref{sec:SystemModel} 
    our formal model comprises four types of elements: 
	(1) signals, (2) PDUs, (3) frames and (4) communication tasks. 
	For each element type we distinguish two types of 
	variables: (1) input parameter, and (2) modeling variables. 
	The input parameter of a task include a timing model  
	for its activation and a deadline. 
	The input parameter for a signal consists of a 
	single timing model and a reference to a PDU. 
	Signal changes are generated by the runnables of the 
	transmitting ECU. For each producing runnable, a timing model 
	describing the points of time the signal value possibly 
	changes is derived from the activation model of the task. 
	These timing models are \emph{summed up} with the help of 
	the $\sqcup$-operation. The resulting timing model is  
	the aforementioned input. %
	The input parameters of a PDU are more diverse. 
	First of all, the different triggering options have to be 
	considered. Besides a direct triggering by contained signals, 
	this can be a threshold for the filling level, and a timeout. 
	Furthermore, for PDUs which are encapsulated in dynamically filled 
	container PDUs, the \emph{collection semantics} are needed to 
	describe possible behaviors. The collection semantics can either 
	be \emph{last-is-best} or \emph{queued}. 
	Queued collection semantics guarantees that every
	instance of the contained PDU is visible 
	on the wire (cf. \cite{Autosar43/IPDUM}).
	Thirdly, the maximum length and the size of the threshold for 
	triggering need to be known to determine a triggering due to 
	the filling level. 
	Finally, a reference to the encapsulating frame is included in 
	the set of input parameters for a PDU.  
	The input parameters of a frame comprise its priority for 
	arbitration, its length and a reference to the communication 
	task responsible for its transmission. These parameters 
	are specific for CAN FD and might need to be adjusted 
	for other physical layer protocols. 

	\begin{table}[tb]
		\centering 
		\caption{Time Variables of Network Model Elements}
		\scriptsize
		\begin{tabular}{l c p{11.6cm}}
			\toprule
			\textbf{Element} & \textbf{Variable} & \textbf{Event description} \\
			\hline
			Signal & $\varphi^S_{i,j}$  & The time span for the $j^\text{th}$ 
			                              change of signal $i$ \\
			PDU    & $\alpha^P_{i,j}$   & The time span in which the $j^\text{th}$ 
			                              instance of PDU $i$ is triggered \\
			PDU    & $\sigma^P_{i,j}$   & The time span in which the $j^\text{th}$ 
			                              instance of PDU $i$ is moved to the lower level buffer \\
			Frame  & $\alpha^F_{i,j}$   & The time span in which the $j^\text{th}$ 
			                              instance of frame $i$ is triggered \\
			Frame  & $\sigma^F_{i,j}$   & The time span in which the $j^\text{th}$ 
			                              instance of frame $i$ is tried to be sent by its sending 
			                              communication task \\   
			Frame  & $\epsilon^F_{i,j}$ & The time span in which the $j^\text{th}$ 
			                              instance of frame $i$ is fully received by its receiving 
			                              communication task  \\
			Task   & $\alpha^T_{i,j}$   & The time span for the activation of the 
			                              $j^\text{th}$ instance of task $i$ \\
			Task   & $\epsilon^T_{i,j}$ & The time span for the completion of the 
			                              $j^\text{th}$ instance of task $i$ \\ 
			\bottomrule	                              
		\end{tabular}
		\normalsize
		\label{tbl:NwVars}
	\end{table}
	
	Besides its input parameters, each model element has different types of 
	modeling variables subjected to the modeling constraints.
	Firstly, there a the \emph{time} related variables listed in Table~\ref{tbl:NwVars}.
	Secondly, each instance of any element is encapsulated in an instance 
	of a lower-layer element. This is modeled in the parameter $n$ for the types 
	signal ($n^S$), PDU ($n^P$) and frame ($n^F$). It contains a reference to the
	instance of the container for each occurrence of the respective element. 
	For frames the semantic is slightly different as $n$ contains the instance of the 
	transmitting communication task in this case.
	Thirdly, in order to obtain safe estimations, a minimum amount of occurrences 
	of each element needs to be considered. Therefore, assume that $\mathbb{T}$ 
	covers a sufficient period of time in which all combinations of relative 
	offsets between occurrences of the timing models appear.  
	Bounds on the length of this period are discussed below. 
	The maximum number of occurrences can be computed for most 
	model elements, if this time interval is fixed. 
	For all the remaining elements, i.e. the container PDUs, 
	the number of occurrences has to be derived from the occurrences of the contained 
	elements. Let $\Omega^S_i = \set{\tOccMin^S_i, \dots, \tOccMax^S_i}$
	be the index set for the occurrences of signal $i$. Let $\Omega^P$, $\Omega^F$, and 
	$\Omega^T$ hold the index set of occurrences for PDUs, frames, and tasks respectively. 
	
	The update of a signal value is modeled 
	with the help of a timing model as described above. Let 
	$t^S_i$ be the timing model of signal $i$. Note that 
	$t^S_i$ can be sporadic or periodic. The $S$ indicates 
	that it belongs to a signal here. 
	The following constraint is added to the model 
	for all $j\in\Omega^S_i$: 
	\begin{align} \label{eq:SignalChanges}
	\varphi^S_{i,j} \geq \pi_1 \left( t^S_i(j) \right) \land 
	\varphi^S_{i,j} \leq \pi_2 \left( t^S_i(j) \right) 
	\end{align}
	
	When a signal value was updated, the so-called 
	\emph{update bit} is set. The update of the value then 
	eventually triggers the sending of a PDU. However, since 
	we are interested in the transmission time for the 
	changed value, we also need to model in which occurrence 
	of its designated container ($\tPDU^S$) the update is transmitted. This 
	is done by adding the following constraints for all signals 
	$i$ and their occurrences $j\in\Omega^S_i$: 
	
	\begin{align}
	\forall k \in \Omega^P_\ell \colon 
		(\varphi^S_{i,j} > \sigma^P_{\ell,k-1} \land \varphi^S_{i,j} \leq \sigma^P_{\ell,k}) 
		\rightarrow (n^S_{i,j} = k)
	\text{ .} 
	\end{align}

	As described above two different events have to 
	be considered for the triggering of PDUs :
	triggering due to timeout 
	modeled by $\alpha^{P-T}$, 
	and 
	triggering due to transmission request by containees 
	modeled by $\alpha^{P-E}$.
	If no clock triggering is configured, 
	$\alpha^{P-T}$ is set to $T_\text{sup}$.
	Analogously, if a PDU is not triggered by 
	any containee, $\alpha^{P-E}$ is set to $T_\text{sup}$.
    The containees of a non-container PDU $i$ are signals. 
    Accordingly, the following constraint is added for all 
    $j\in\Omega^P_i$:
	\begin{align} 
	\alpha^{P-E}_{i,j} = \min\set{\varphi^S_{\ell,k} | \ell\in \tSignals^P_i,k\in \Omega^S_\ell \land n^S_{\ell,k} = j} 
	\text{.}
	\end{align}
	
	For container PDUs, possibilities for triggering are more complex. 
	The following points of time are considered conditionally: 
	the point of time the first containee was triggered 
	$\alpha^{P-C1}_{i,j}$, 
	the point of time the first containee was triggered 
	plus the timeout of the container $\alpha^{P-CT}_{i,j}$, and
	the point of time the length of the contained PDUs 
	exceeds the threshold of the container $\alpha^{P-Cn}_{i,j}$. 
	To detect a triggering of a container PDU due to exceeding 
	of the threshold, the filling level needs to be determined. 
	To this end, we introduce an additional 
	variable foreach instance $j$ of a PDU $i$, $len^P_{i,j}$, which contains 
	the total length of the PDU. Additionally, for a pair of 
	PDUs $i$ and $j$ and each instance $j$ of $i$ and $k$ of $\ell$ 
	we add an auxiliary variable $c^P_{i,j,\ell,k}$ which holds 
	$1$ if $k$ is contained in $j$ and $0$ otherwise, 
	i.e. 
	\begin{align}
	c^P_{i,j,\ell,k} = 
	\begin{cases}
	1 & \text{ if } \ell\in\tPDU^P_i \land n^P_{\ell,k} = j  \\
	0 & \text{ else } 
	\end{cases}
	\end{align}
	for all PDUs $i,\ell$ and $j\in\Omega^P_i$, $k\in\Omega^L_i$. 
	
	Depending on the collection semantics, an instance of 
	a PDU might be overwritten if its container is not 
	send between two updates. This means, that 
	$c^P$ is $1$ for two instances of the contained 
	PDU. To model the fact that an instance $k$ of a
	PDU $\ell$ 
	might be overwritten by a subsequent instance in 
	the instance $j$ of its container PDU $i$, 
	we add an 
	additional variable $o^P_{i,j,\ell,k}$ 
	with 
	\begin{align}
	o^P_{i,j,\ell,k} = 
	\begin{cases}
	1 & \text{ if } c^P_{i,j,\ell,k} = 1 \land \exists k'\in\Omega^P_\ell \colon k' > k \land n^P_{\ell,k'} = j  \\
	0 & \text{ else } 
	\end{cases}
	\end{align}
	for all PDUs $i,\ell$ and $j\in\Omega^P_i$, $k\in\Omega^P_\ell$. 
		
	The length of a non-container PDU is fixed, based on the 
	contained signals and a fixed-size header. 
	The length of a container PDU depends on its PDU 
	layout. If it has a \emph{static} layout, the length 
	is fixed. Otherwise, if it has a \emph{dynamic} layout, 
	the collection semantics of the contained 
	PDUs is the crucial factor. 
	If the collection semantic is \emph{last-is-best} 
	the content in the container can be overwritten.
	Otherwise, if the collection semantic is 
	\emph{queued}, multiple instances of the same 
	PDU can be transmitted in one container. 
	Note, that we assume that containers cannot be 
	nested.  
	The actual length of a container can therefore 
	finally be calculated by summing the length of 
	all not-overwritten containee instances, i.e. 	
	\begin{align} \label{eq:LengthContainer}
	len^P_{i,j} = \tLength^{P-\text{H}}_{i} + \sum_{\substack{\ell\in \tPDU^P_i,\\ k\in \Omega^P_\ell}} 
		(1- o^P_{i,j,\ell,k}) \cdot c^P_{i,j,\ell,k} \cdot len^P_{i,j} 
	\end{align} 
	for all PDUs $i,\ell$ and $j\in\Omega^P_i$, $k\in\Omega^P_\ell$ 
	where $\tLength^{P-\text{H}}_{i}$ is the length of the header.  
	In order to ensure correct modeling of the 
	collection semantics, the following constraints 
	are added conditionally: 
	\begin{align}
	 n_{i,j} &\leq n_{i,{j+1}} &\text{if } i \text{ is collected \emph{last-is-best}} \\
	 n_{i,j} &< n_{i,{j+1}}    &\text{if } i \text{ is collected \emph{queued}}
	\end{align}

	To determine the possible point of time for the triggering 
	of a container PDU, the minimum of the values is 
	used: 
	\begin{align}
	\alpha^P_{i,j} = \min\set{ \alpha^{P-C1}_{i,j}, \alpha^{P-CT}_{i,j}, \alpha^{P-Cn}_{i,j} } \text{.}
	\end{align}
	If one of the triggers is not applicable, the respective 
	value is set to $T_\text{sup}$.   
	This means, if $\alpha^P_{i,j} = T_\text{sup}$ the instance of the container PDU 
	has not been triggered and must not be considered further if not triggered due to 
	a timeout.  
	Finally, the triggering of an instance $j$ of an PDU $i$ happens 
	at the first point of time it is possibly triggered by any 
	of the described triggers, i.e. 
	\begin{align} 
	\alpha^P_{i,j} = \min\set{ \alpha^{P-E}_{i,j}, \alpha^{P-T}_{i,j} } 
	\end{align}
	for all PDUs $i$ and PDU occurrences $j\in\Omega^P_i$.  
		
	Given $\alpha^P_{i,j}$ it can be described in which occurrences of the 
	encapsulating PDU ($\tPDU^P$) $j$ is possibly transmitted. However, 
	since we are considering container PDUs with dynamic layouts
	decided at runtime, a PDU is not necessarily send within 
	the next instance of a container. In other words, if the 
	container PDU is already filled to capacity, the 
	containee has to wait until the next instance 
	is sent. To model this, we constrain all instances 
	of the container which are sent between the instance 
	encapsulating the an instance $j$ of an PDU $i$ 
	and the point of time $j$ was triggered to be too 
	full to contain $j$. The variable 
	$n^P_{i,j}$ holds index of the encapsulating instance 
	for all PDUs $i$ and $j\in\Omega^P_i$. 
	The variable $f\!n^P_{i,j}$ holds the index of the first instance 
	which is a candidate for encapsulation, i.e.
	for all PDUs $i$ and PDU occurrences $j\in\Omega^P_i$ which 
	are mapped to a container PDU $\ell$, 
	\begin{align}
	\begin{split}
	f\!n^P_{i,j} = 
		\min
		\left(
		\set{\tOccMax^P_\ell} \cup 
		\set{ k | k \in \Omega^P_\ell \land \alpha^P_{\ell,k} \geq \alpha^P_{i,j} }
		\right) 
	\end{split} \text{ .} 
	\end{align}
		
	If the collection semantic of the PDU $i$ 
	into the container PDU $\ell$ is \emph{queued}, 
	the following constraint is added for each occurrence 
	$j$ in $j\in\Omega^P_i$:
	\begin{align}
	\begin{split} \label{eq:PDUTransmission0}
	\forall k\in\Omega^P_\ell \colon 
		(k \geq f\!n^P_{i,j} \land k > n^P_{i,j-1} \land k < n^P_{i,j}]) 
		\rightarrow 
		(len^P_{\ell,k} + len^P_{i,j} > \tLength^P_\ell)
	\end{split} \text{.} 
	\end{align} 
	Otherwise, if the collection semantic of the PDU $i$ into 
	the container PDU $\ell$ is 
	\emph{last-is-best}, the constraints for $j$ in $j\in\Omega^P_i$
	are depending on whether an instance of the same PDU 
	is already part of the container. If this is the case, 
	the content would simply be overwritten. Otherwise, 
	if there exists no $j'\in\Omega^P_i\setminus\set{j}$
	such that $n^P_{i,j'} = k$, the following constraint 
	needs to be added:  
	\begin{align}
	\begin{split} \label{eq:PDUTransmission1}
	\forall k\in\Omega^P_\ell \colon 
		(k \geq f\!n^P_{i,j} \land k < n^P_{i,j}) 
		\rightarrow 	
		(len^P_{\ell,k} + len^P_{i,j} > \tLength^P_\ell)
	\end{split} \text{.} 
	\end{align} 
	It is important to note that in this case, instances 
	of PDU $i$ must be excluded when computing the 
	length of $\ell$ as previously contained instances 
	would be overwritten. 
	
	Together with the constraint for the length of a 
	container PDU, the Constraints~\ref{eq:PDUTransmission0} 
	and \ref{eq:PDUTransmission1} assure that if 
	no non-full container can be sent if one of its 
	contained PDU is queued for sending and could fit 
	into the container instance lengthwise. 
		
	Once triggered, the frame is 
	queued for transmission on the bus.
	Thereupon, the frame data is copied from global 
	memory to the buffer of the communication 
	controller by the next instance of the 
	responsible communication task. 
	Therefore, in order to get the relative temporal distance 
	between sending and receiving communication task, the possible point of 
	time it was queued for transmission needs to be tracked for 
	an instance of a frame. 
	The following constraint is added for all frames $i$ and 
	their instances $j\in\Omega^F_i$: 
	\begin{align}
		\alpha^F_{i,j} = \min\set{\alpha^P_{\ell,k} | \ell\in \tPDU^F_i, k\in \Omega^P_\ell \land n^P_{\ell,k} = j }
		\text{.} 
	\end{align}
 	
	In the following, we describe the constraints needed to model 
	CAN FD network. They can easily be plugged in by adding the 
	corresponding constraints. The constraints listed up to this point 
	can consequently be reused for other bus types. 
	Let $iTx$ be the communication task for frame 
	$i$ and $jTx$ be the first instance of this $iTx$ after 
	occurrence $j$ of $i$ was queued for transmission. Furthermore, 
	let $\alpha^T_{iTx,jTx}$ be the activation time of this task and
	$\tFrame^{HP}_i$ the set of frames which have a higher priority 
	than $i$. The following constraint is added to model the 
	start of transmission: 
	\begin{align}
	\begin{split} \label{eq:SigmaFrame}
		\sigma^F_{i,j} &= \max\set{ \alpha^T_{iTx,jTx} } \cup 
		 \set{ \epsilon^F_{\ell,k} | \ell\in \tFrame^{HP}_i, k\in \Omega^F_\ell \land \sigma^F_{\ell,k} \leq \sigma^F_{i,j} }
		\text{.}
	\end{split} 
	\end{align}
	
	For CAN FD, the transmission time of a frame is
	resulting from an arbitration phase and the  
	time the bits are transmitted on the physical medium. In the 
	arbitration phase, all nodes of 
	the network agree on the node allowed to transmit next. 
	The node trying to transmit the frame with the highest 
	priority is allowed to continue sending after arbitration. 
	From Constraint~\ref{eq:SigmaFrame} it can be inferred 
	that the occurrence $j$ of frame $i$ has the highest 
	priority at point of time $\sigma^F_{i,j}$. Therefore, 
	the end of transmission $\epsilon^F_{i,j}$ can be calculated 
	in the following way: 
	\begin{align}
		\left\lfloor t_\text{arb} + (len^F_{i,j} \cdot t_\text{bit}) \right\rfloor \leq \epsilon^F_{i,j} \leq  
		\left\lceil t_\text{arb} + (len^F_{i,j} \cdot t_\text{bit}) \right\rceil
	\end{align} 
	where $t_\text{arb}$ worst-case time needed for arbitration and 
	$t_\text{bit}$ is the time needed to transfer one bit of data. 
	The length of the instance $j$ of frame $i$, $len^F_{i,j}$, 
	can be computed analogously to the PDUs 
	(cf. Constraint \ref{eq:LengthContainer}). 	
	With the help of $\epsilon^F_{i,j}$ the instance of the 
	communication task at the receiving ECU can be determined. 
	The time between the initial change of the signal and the 
	deadline of this task is the time needed for a value change of 
	this signal to be transferred from the global memory of one 
	ECU to the next. 
	Depending on whether the clocks of the ECUs in the network are 
	synchronized, a clock drift can additionally be considered for 
	the receiving communication task.  
	
	Given all constraints for a set of network artifacts, 
	a constraint solver is deployed to obtain the worst-case transmission 
	time. The solver  searches all satisfying assignments 
	for the one modeling the longest transmission time. 
	This emulates an exhaustive search if all situations are covered. 
	To guarantee this, all possible relative offset between the 
	PDU containing the objective signal and all
	PDUs possibly causing a delay of this PDU needs to be 
	covered. If all relevant PDUs are triggered by 
	periodic timers or periodically changing signals, 
	the least common multiple of these periods covers 
	all relative offsets. If sporadic events are included, 
	a lot more relative offsets are possible. However, 
	in this case the solver freely chooses values between the 
	first and last possible point of time for the event 
	to happen (cf. Constraint~\ref{eq:SignalChanges}).
	For practical application, it is furthermore 
	important that $\mathbb{T}$ covers a sufficiently large time
	interval such that it is possible that all PDUs which are triggered 
	can also be sent.  
	
	\subsection{Estimations for Software Task Chains} \label{sec:EstimationsSW}
	Software task chains can either start with a sensor reading or a 
	network communication task and end at an actuator or 
	another network communication task. Regardless of type, 
	estimating latencies for task chains is concerned with the question 
	which relative offsets are possible between task instances on the chain. 
	The activation pattern of the task, the method for accessing memory, 
	as well as the core execution times are 
	factors affecting the possibilities for these relative offsets. 
	In Section~\ref{sec:RelatedWork} state-of-the art approaches to 
	obtain latencies for different kind of task chains are listed. 
	We want to highlight the approach described in 
	\cite{DBLP:conf/sies/FrieseEN18} 
	as it provides the interesting possibility to create a combined 
	model for both types of sub-chains considered in this work. 
	The combination of both approaches would allow for an end-to-end 
	timing model comprising multiple ECUs and networks. 
	Although attention needs to be payed to potential scalability problems 
	such a model is promising in terms of the precision of the results. 
	
	\subsection{End-to-end Estimations} \label{sec:EstmationsE2E}
	Due to complexity, estimations of end-to-end latencies are often 
	performed on smaller, easier to handle segments of a chain. 
	This comes with the price of a \emph{context loss} at each 
	\emph{cut} of the analysis.  After a context loss, most information 
	on the situation which lead to the worst case in previous segment 
	need to be dropped. 
	As a result, mutually exclusive situations might be 
	considered cumulatively. This can lead to significant overestimations.
	To reduce the impact of context losses, 
	we propose to cut analysis only at communication tasks.  
	This cut essentially results in two kind of sub chains: 
	communication task to communication task via software 
	and via network. 
	An advantage of cutting analysis at communication tasks is, 
	that context losses with regard to changes of signal 
	values are not too costly in terms of information loss. 
	The task chain on the next ECU certainly starts with 
	the receiving communication task where local signals of the 
	sending ECU have no direct impact. Moreover, having a context loss 
	prior to determining the possibilities for the PDU trigger 
	is unavoidable with many approaches to estimate 
	task chains. They only keep track of the one signal relevant 
	for the chain. However, as described above, different signals
	can be decisive for the triggering of the PDU transporting the 
	signal over the network. With the approach presented in this work, 
	the trigger mechanisms for PDUs can be considered 
	independently.	

	\section{Experiments} \label{sec:Experiments}
		
	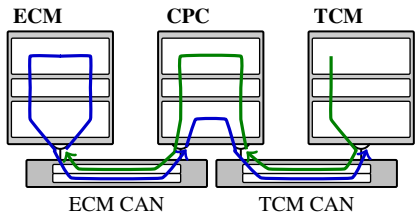
\begin{wrapfigure}[11]{l}{6.2cm}
		\centering 
		
		\scriptsize
		\begin{tikzpicture}[xscale=0.5,yscale=0.6] 
		\tikzstyle{ecu} = [line width=0.75pt, color=black, fill=black!20!white, rounded corners=0.5pt]
		\tikzstyle{con} = [line width=0.75pt, color=black]
		\tikzstyle{bus} = [line width=0.75pt, color=black, fill=black!20!white, rounded corners=0.5pt]
		\tikzstyle{sw}  = [line width=0.50pt, color=black, fill=white, rounded corners=0.5pt]
		\tikzstyle{art} = [line width=0.50pt, color=black, fill=white, rounded corners=0.5pt]
		\tikzstyle{chainA} = [->, color=green!50!black, line width=1.2pt, smooth]
		\tikzstyle{chainB} = [->, color=blue!80!black, line width=1.2pt, smooth]
		
		\foreach \x/\n in { 0.0/ECM, 4.0/CPC, 8.0/TCM} { 
			\draw[ecu] (\x+ 0.00,-0.05) rectangle ++ ( 2.80, 2.50) ; \node[draw=none] at (\x+ 0.80,2.80) {\textbf{\n}};
			\draw[sw ] (\x+ 0.10, 0.10) rectangle ++ ( 2.60, 0.80) node[draw=none,pos=.5] (BSW\n) {}; 
			\draw[sw ] (\x+ 0.10, 1.00) rectangle ++ ( 2.60, 0.40) node[draw=none,pos=.5] (RTE\n) {}; 
			\draw[sw ] (\x+ 0.10, 1.50) rectangle ++ ( 2.60, 0.80) node[draw=none,pos=.5] (ASW\n) {}; 
		}
				
		\draw[bus,fill=black!25] ( 0.50,-1.00) rectangle ++ ( 4.80, 0.60) node[midway, below=.2] {ECM CAN};
		\draw[art]     ( 1.20,-0.90) rectangle ++ ( 3.40, 0.20) node[draw=none,pos=.5] (PDU2) {};
		\draw[art]     ( 1.20,-0.70) rectangle ++ ( 3.40, 0.20) node[draw=none,pos=.5] (PDU1) {};
		\draw[con, -(] ( 1.40, -0.40) -- ++( 0.00, 0.37) node[draw=none,pos=.5] (COMECM) {};
		\draw[con, -(] ( 4.60, -0.40) -- ++( 0.00, 0.37) node[draw=none,pos=.5] (COMCPC0) {};

		\draw[bus,fill=black!25] ( 5.55,-1.00) rectangle ++ ( 4.80, 0.60) node[midway, below=.2] {TCM CAN};	
		\draw[art]     ( 6.25,-0.90) rectangle ++ ( 3.40, 0.20) node[draw=none,pos=.5] (PDU3) {};
		\draw[art]     ( 6.25,-0.70) rectangle ++ ( 3.40, 0.20) node[draw=none,pos=.5] (PDU0) {};
		\draw[con, -(] ( 6.20, -0.40) -- ++( 0.00, 0.37) node[draw=none,pos=.5] (COMCPC1) {};
		\draw[con, -(] ( 9.40, -0.40) -- ++( 0.00, 0.37) node[draw=none,pos=.5] (COMTCM) {};
		
		\draw [chainA] plot [smooth, tension=.15] coordinates { 
			($ (ASWTCM)   + (-0.80, 0.00) $) 
			($ (RTETCM)   + (-0.80, 0.00) $) 
			($ (BSWTCM)   + (-0.80, 0.00) $) 
			($ (COMTCM)   + (-0.15, 0.05) $)
			($ (COMTCM)   + (-0.15, 0.00) $)  
			($ (PDU0)     + ( 1.00, 0.00) $) 
			($ (PDU0)     + (-1.00, 0.00) $)
			($ (COMCPC1)  + ( 0.15, 0.00) $) };
		
		\draw [chainA] plot [smooth, tension=.15] coordinates { 
			($ (COMCPC1) + ( 0.15, 0.05) $)
			($ (BSWCPC)  + ( 0.80, 0.00) $) 
			($ (RTECPC)  + ( 0.80, 0.00) $) 
			($ (ASWCPC)  + ( 0.80, 0.00) $)   
			($ (ASWCPC)  + (-0.80, 0.00) $) 
			($ (RTECPC)  + (-0.80, 0.00) $) 
			($ (BSWCPC)  + (-0.80, 0.00) $)
			($ (COMCPC0) + (-0.15, 0.05) $)   
			($ (COMCPC0)  + (-0.15, 0.00) $)  
			($ (PDU1)     + ( 1.00, 0.00) $) 
			($ (PDU1)     + (-1.00, 0.00) $)
			($ (COMECM)   + ( 0.15, 0.00) $) };
		
		\draw [chainB] plot [smooth, tension=.15] coordinates { 
			($ (COMECM)   + ( 0.15, 0.05) $)
			($ (BSWECM)   + ( 0.80, 0.00) $) 
			($ (RTEECM)   + ( 0.80, 0.00) $) 
			($ (ASWECM)   + ( 0.80, 0.00) $)   
			($ (ASWECM)   + ( 0.00, 0.00) $) 
			($ (ASWECM)   + ( 0.00, 0.00) $)
			($ (ASWECM)   + (-0.80, 0.00) $)
			($ (RTEECM)   + (-0.80, 0.00) $) 
			($ (BSWECM)   + (-0.80, 0.00) $)
			($ (COMECM)   + (-0.15, 0.05) $)
			($ (COMECM)   + (-0.15, 0.00) $)
			($ (PDU2)     + (-1.20, 0.00) $) 
			($ (PDU2)     + ( 1.20, 0.00) $)
			($ (COMCPC0)  + ( 0.15, 0.05) $)  }; 
		
		\draw [chainB] plot [smooth, tension=.15] coordinates { 
			($ (COMCPC0) + ( 0.15, 0.10) $)		
			($ (BSWCPC)  + (-0.40, 0.00) $) 
			($ (BSWCPC)  + ( 0.40, 0.00) $)
			($ (COMCPC1) + (-0.15, 0.05) $) 
			($ (COMCPC1) + (-0.15, 0.00) $)  
			($ (PDU3)     + (-1.20, 0.00) $) 
			($ (PDU3)     + ( 1.20, 0.00) $)
			($ (COMTCM)   + ( 0.15, 0.05) $) };
		  		
		\end{tikzpicture}
		\caption{Case-Study: Topology and Chain}
		\label{fig:CaseStudy}
	\end{wrapfigure}
	
	To test the constraint model for the estimation of
	signal transmission times, we applied it on a case study 
	based on real automotive data. 
	It contains synthetic but realistic data for two 
	CAN-FD networks connecting three ECUs 
	as depicted in Figure~\ref{fig:CaseStudy}. 
	The amount of frames and PDUs roughly matches 
	the amount of artifacts for the functional 
	communication between the powertrain domain 
	controller (CPC), the engine controller (ECM), 
	and the transmission controller (TCM). 	
	Responses to diagnostic messages as well as 
	network management communication is not 
	considered. 
	Furthermore, only signals 
	potentially having an impact on the triggering 
	of a PDU are included. Two signals are 
	aggregated to one signal if it has no impact 
	on the estimation. More precisely, two signals 
	are combined to one modeling the same behavior 
	if they are sent as part of the same PDU with 
	the same trigger-related properties and periodic 
	timings and neither is the objective signals.  
	In this way, the amount of signals in the model is 
	greatly reduced when compared to a naive model. 
			
	We analyzed the worst-case time for the transmission 
	of eight different signals, four contained in 
	dynamic container PDUs, four being part of the cause-effect 
	chain depicted in Figure~\ref{fig:CaseStudy}.
	The first-mentioned signals represent status 
	updates sent from the ECM and TCM to the CPC. 
	The cause-effect chain on the other hand is 
	a control loop and has two parts.
	The first part of the chain describes the data flow when 
	the TCM sends a request for torque to the ECM. 
	The CPC checks the requests before forwarding it to the ECM. 
	After providing torque to the extend possible, the ECM reports 
	to the CPC. The answer of the ECM is directly routed from the 
	ECM CAN to the TCM CAN for fast reception at the TCM. 
	
	To implement the constraint model described in 
	Section~\ref{sec:FormalAbstractions} we used the 
	high-level constraint language MiniZinc. The 
	constraints can be translated almost directly. 
	An advantage of using MiniZinc 
	is the descriptive nature of the resulting model. 
	It is readable and verifiable for anyone with 
	a basic understanding of constraint modeling. 
	The greatest benefit however is the form of the results. 
	If a feasible solution was found, constraint solvers 
	do not only answer \emph{yes} or \emph{no} but 
	they give a assignment for all variables of the 
	constraint set. 
	The feasible solutions for our set of time and 
	decision variables can directly be interpreted 
	to identify the situation which lead to the worst 
	case. 
	Each set of artifacts is encoded in a data file to replace 
	constraint parameters with their actual values. This 
	data file is then linked to the constraints and 
	compiled to a second constraint model in the 
	low-level constraint language FlatZinc. FlatZinc is 
	supported by different solver back-ends. 
	We used the parallel version of the lazy-clause generation 
	constraint solver \emph{Chuffed} \cite{DBLP:conf/cpaior/EhlersS16}.
	The experiments have been carried out on a  desktop computer 
	equipped with an Intel(R) Core(TM) i9-7940X CPU and 
	128GB of memory. 
	The time and memory needed to compile the MiniZinc model as well as  
	the time and memory needed for solving the FlatZinc model by 
	16 instances of \emph{Chuffed} working in parallel 
	are shown in Table~\ref{tbl:Performance}. 
	
	\newcommand{\TCA}{${}^{\mathbf{*}}$}
	\begin{table}[tb]
		\centering 
		\caption{Resource Usage for Compiling and Solving the Model}
		\begin{tabular}{c c c r r r r r c r}
		\toprule
		\multicolumn{1}{p{1.20cm}}{{\textbf{Sender}}} & 
		\multicolumn{1}{p{1.20cm}}{{\textbf{Receiver}}} & 
		\multicolumn{1}{p{2.20cm}}{{\textbf{Signal}}} & 
		\multicolumn{2}{p{2.20cm}}{\centering{\textbf{Compiler}}} & 
		\multicolumn{1}{p{0.20cm}}{\centering{                 }} & 
		\multicolumn{2}{p{2.20cm}}{\centering{\textbf{Solver}}}   & &
		\multicolumn{1}{p{1.50cm}}{\centering{\textbf{Result}}}  \\
		    &          &                       &       s &   kbyte &&      s  &  kbyte  & & $\mu$s \\ 
		\cmidrule(l{.75em}){4-5} \cmidrule(l{1.75em}){6-8} \cmidrule(l{1.75em}){10-10}
		
		ECM & CPC  & Status~A    &    2:29 &   4,435,568 &&    2:30  &   57,462,088 &\TCA &   15500 \\  
		ECM & CPC  & Status~B    &    2:28 &   4,441,504 &&    2:16  &   60,797,024 &\TCA &   20500 \\  
		TCM & CPC  & Status~C    &    0:55 &   1,723,920 &&    0:46  &    8,772,148 &     &   15500 \\  
		TCM & CPC  & Status~D    &    0:52 &   1,723,896 &&    1:03  &    9,863,388 &     &   15500 \\  
		TCM & CPC  & Request~1   &    0:52 &   1,723,808 &&    0:41  &    8,499,072 &     &   11250 \\  
		CPC & ECM  & Request~2   &    2:24 &   4,432,880 &&    3:02  &   19,298,896 &     &   12500 \\  
		ECM & CPC  & Response~1  &    2:27 &   4,466,500 &&    2:45  &   91,870,500 &     &    5500 \\  
		CPC & TCM  & Response~2  &    0:56 &   1,734,188 &&    0:49  &    9,050,628 &     &    6000 \\  
		\bottomrule
		\end{tabular}	
		\scriptsize
		\begin{flushright}
	    \TCA Running instances of the solver were interrupted after the optimum was found.  
		\end{flushright}
		\normalsize	
		\label{tbl:Performance}
	\end{table}
	
	\section{Future Work} \label{sec:FutureWork}
	In future work we want to examine the possibilities 
	arising from the integration of software-level and 
	network-level models. The scalability of a holistic 
	model must be examined. If it can be contained 
	by with current hardware and solver technologies, 
	it could not only be used to verify 
	system designs but also to automatically generate 
	suggestions on the optimization. 
	In distributed system with cause-effect chains 
	spanning over multiple ECUs the effects of changes 
	are very difficult to predict and might influence 
	the performance of other cause-effect chains. 
	With precise end-to-end estimations however, these 
	impacts could be estimated quickly to evaluate
	system designs quickly. 
	
	\section{Conclusion} \label{sec:Conclusion}
	In this paper we addressed the problem of estimating 
	the end-to-end latency of distributed cause-effect 
	chains in automotive systems.
    A previously unsupported part of the problem is estimating the 
    temporal behavior of the PDU triggering mechanism.  
    For this, we introduced timing models which 
	can be combined to model the temporal behavior of 
	two or more event sources. Based on this we presented 
	a constraint model to estimate the time needed to 
	transmit signal changes in CAN FD communication 
	clusters. We discussed how this approach can be 
	integrated in an end-to-end analysis and why this 
	would improve estimations. Finally, the application 
	on an OEM use case shows scalability for industrial-scale 
	problems. 
	
	\bibliographystyle{eptcs}
	\bibliography{paper06}
	
\end{document}